IAC-23-E2.2.5

# Low-cost, Lightweight Electronic Flow Regulators for Throttling Liquid Rocket Engines

**Vint Lee[a]\*, Sohom Roy[a]**

[a] *UC Berkeley, 110 Sproul Hall Berkeley, CA 94720*
\* Corresponding Author (vint@berkeley.edu)

**Abstract**

For small-scale liquid rockets, pressure-fed systems are commonly favoured due to their simplicity and low weight. In such systems, accurate regulation of both tank and injector pressures over a wide range of upstream pressures is critical — more accurate regulation allows for higher engine efficiency and minimal tank mass, thus improving flight performance. However, existing methods such as dome-loaded pressure regulators are inflexible, or require extensive characterization to function accurately. These methods also suffer from limited orifice size, droop, and slow reaction times, making them unsuitable for throttling by adjusting pressures in flight, which are increasingly important as propulsively landing rockets become more common. To overcome these challenges, we designed an electronic pressure regulator (eReg), a multi-input multi-output system utilising closed loop feedback to accurately control downstream pressures. Our design is simple, low-cost and robust: with a single ball valve actuated by a motor, we regulate both gaseous pressurant and cryogenic liquid propellant at high flow rates (1.14 kg/s of liquid; 0.39 kg/s of gas) and upstream pressures (310 bar). Using 2 eRegs to regulate propellant tank pressures, and 2 eRegs for regulating propellant flow to the engine, we demonstrated our system's ability, in a static fire test, to regulate pressures accurately (within 0.5 bar) while simultaneously throttling our engine. To the best of our knowledge, this is the first time any undergraduate team has successfully throttled a liquid bipropellant engine.

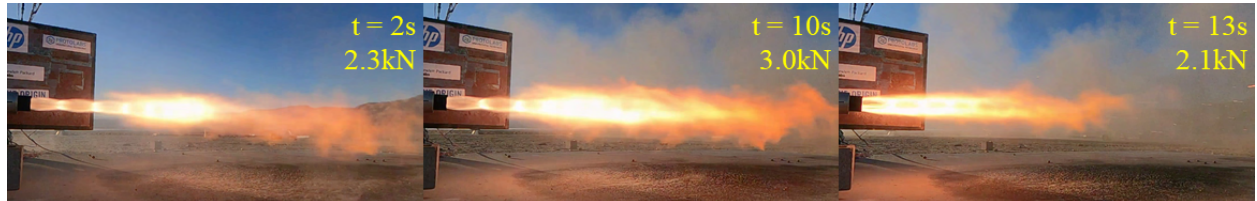

**Fig. 1.** Static fire test demonstrating our electronic regulators' (eReg) ability to throttle up and down.

**Acronyms/Abbreviations**
MEOP - Maximum expected operating pressure
PT - Pressure transducer
PID - Proportional-integral-derivative

## 1. Introduction

The need for robust pressure-reducing regulators that are able to operate throughout large changes in upstream pressure is becoming increasingly common as more work is done on small-scale pressure-fed liquid rocketry. Accurate regulation of both tank pressures and injector pressures can allow for higher engine efficiency and optimised tank construction with the system operating closer to target design parameters.

Methods currently used for gas regulation, including dome-loaded pressure regulators and solenoid systems, are often inflexible and require extensive characterization [6, 8]. Moreover, electro-mechanical limitations such as limited orifice size, droop, slow reaction times, and cryogenic incompatibility [5] make them unsuitable for applications such as throttling, which are increasingly important as vertical take-off vertical landing rockets become more ubiquitous.

To overcome these challenges, we designed an electronic pressure regulator (eReg), a multi-input multi-output system utilising closed loop feedback to accurately control downstream pressures. Consisting of a single ball valve actuated by a motor, its robust design allows eRegs to regulate both gaseous pressurant and cryogenic liquid propellant at high flow rates (1.14 kg/s of liquid; 0.39 kg/s of gas) and upstream pressures (310 bar). By using 4 eRegs in conjunction (2 regulating propellant tank pressures and the other 2 regulating propellant flow to the engine), we are able to operate our engine at a near optimal oxidizer-fuel ratio while throttling over a wide range of thrust.

To reduce system mass, powerful motors and compact gearboxes were used. For high flow rates,

serviceability, and manufacturing considerations, commercial full port ball valves were chosen as the flow control orifice. High flight reliability in the face of vibrations was accomplished through the use of castellated components and thread-locking components.

Each eReg uses a cascaded controller, with the outer PID loop monitoring downstream pressure and computing valve angle setpoints for the inner PID loop. However, the control problem is complicated by the constantly changing ullage volume and upstream pressures, which result in a non-linear, time-dependent system. We found that adding feedforward and using dynamic PID gains significantly improved performance, thus allowing us to achieve precise control despite the challenging environment.

We also verified the design's performance and reliability through a testing campaign, culminating in a static fire test demonstrating eReg's ability to regulate pressures accurately (within 0.5 bar) while simultaneously throttling our engine. Images from this test are shown in Figure 1. To the best of our knowledge, this is the first time any undergraduate team has successfully throttled a liquid bipropellant engine.

## 2. Material and methods
*2.1 System Design and P&ID*

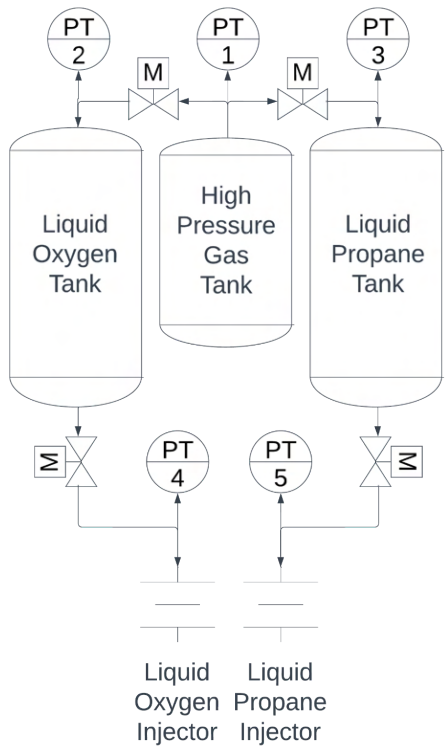

**Fig. 2.** P&ID of the test stand.

The test cell utilised for pressure regulation testing was a liquid bipropellant engine test stand developed by the Space Enterprise at Berkeley undergraduate rocketry team. The test stand was designed for cryogenic operation with liquid oxygen as the oxidiser and liquid propane or methane as the fuel. A commercially available type 3 composite overwrapped pressure vessel typically used for firefighting applications was pressurised with gaseous nitrogen to pressurise the fuel and oxidiser tanks to a set pressure. The propellant tanks were seamless aluminium tanks ported to accommodate the outflow of propellant, inflow of pressurisation gas, and additional system hardware.

The P&ID (Figure 2) elaborates on the location of regulating valves and other components on the system. Valves placed upstream of the propellant tanks regulated the tank pressure to a constant level by controlling the entrance of nitrogen gas pressurant. Valves placed downstream of the propellant tanks regulated injector pressures by changing the pressure drop across the valve orifice throughout the flow. Not shown in the P&ID were relief valves and burst disks for system safety in the event of an overpressurization event, and check valves used to accommodate propellant fill and prevent propellant entering the pressurant tank.

The thrust chamber used during testing was an ablative chamber tested to durations of 14 seconds. At 100% design thrust the oxidiser mass flow rate was 1.14kg/s and the fuel mass flow rate was 0.49kg/s. The nominal thrust was 3 kN with a chamber pressure of 24 bar. The thrust chamber was externally ignited with solid propellant. The test cell was manually loaded with propellants and remotely operated during pressurisation and flows.

*2.2 Valve Selection*

Valve selection was primarily performed by considering pressure loss across the valve and the availability of suitable commercial valves. Initial sizing was based on maximum future flexibility for the existing test stand by matching the valves maximum orifice size to the orifice size of the plumbing used on the stand. Thus, full port ball valves were chosen for their ability to permit maximum flow at minimal size and short throw distances [2]. In the case of an upsizing of the feed system mass flow rate requirements, the hardware likely would have continued to be appropriate.

Valve selection for the gaseous pressure regulation function was chosen with a rated pressure of 415 bar. Maximum Expected Operating Pressure for them was 310 bar. Valves selected for propellant injector pressure regulation were chosen with a rated pressure of 78 bar and operated at a MEOP of 42 bar.

*2.3 Hardware Design*

After suitable valves were selected for regulation, hardware was designed to rapidly actuate them. Brushed motors were chosen initially for their ease of operation during much of testing, though brushless motors were chosen for a flight vehicle for ease of packaging. Figure 3 shows the hardware design of a single eReg unit using the brushed motors, while Figure 4 shows a picture of two eRegs built and mounted on the test stand.

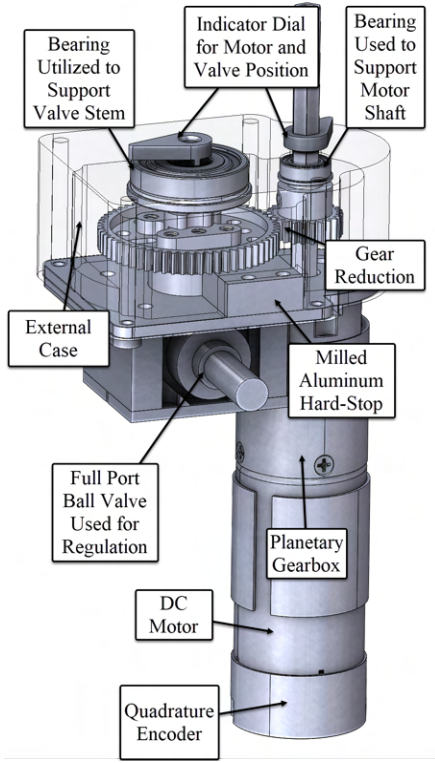

**Fig. 3.** eReg hardware components.

At a component level design, multiple decisions were made to increase the reliability of hardware.

An aluminium hub was pressed onto the stainless steel valve shaft, allowing for minimal backlash and robust performance in environments with heavy vibrations, where set screws or other locking features may have failed. The motor was face mounted to a robust planetary gearbox allowing for a compact gear reduction that was resistant to shock loading that would damage spur gearboxes with a similar form factor. For packaging and a further reduction, a spur gear was bolted to the valve hub. The output of the planetary gearbox was a hex shape, and the gear was broached with a hex shape, allowing for a positive drive.

The nominal gear spacing of the spur gear reduction was adjusted to reduce backlash. The reduction in gear life was deemed acceptable. For feedback, an incremental encoder was placed directly on the rear shaft of the motor. The planetary and spur gear reduction allowed a virtual increase in encoder resolution. To zero the encoder, the motor was stalled against a hard-stop, though in the flight vehicle design limit switches were added as the higher power motor chosen damaged the mechanical hard-stops when stalled. The hard-stops and limit switches also limited the travel to 90 degrees, the full throw of the ball valves. To prevent excessive side loading on the planetary gearbox shaft and valve stem during rapid actuations, ball bearings were used to support the shafts.

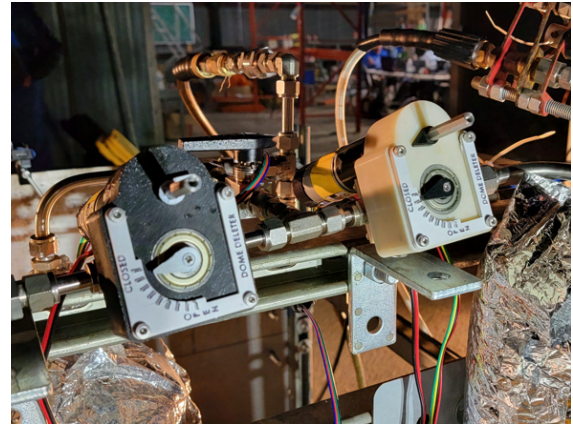

**Fig. 4.** Tank eRegs built and mounted.

*2.4 Controller Design*

To achieve high regulation accuracy, and for flexibility in adjusting pressure setpoints, we used a feedback control system, with the controller implemented in software running on ESP32 microcontrollers. This design not only allows us to iterate and tune the controller rapidly, but also allows us to throttle the engine during flight by simply changing the pressure setpoints in software. For simplicity, each of the four eRegs (tank and injector for both propellants) used an independent controller running on separate boards, with telemetry and commands sent via ethernet. The rest of this section details the design of an individual controller.

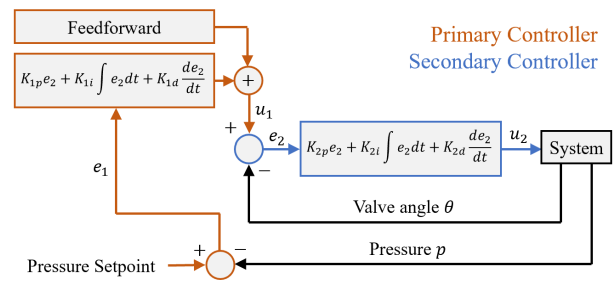

**Fig. 5.** eReg uses a cascaded controller architecture.

A cascaded controller architecture was used, with the primary controller used to set the target valve position in response to readings from a downstream pressure transducer (PT), and the secondary controller to drive the motor to reach the desired valve position. A diagram of the controller architecture is shown in Figure 5, with the same architecture being used in both tank eRegs and injector eRegs. This architecture was primarily chosen as the motor dynamics were significantly faster than the pressurant dynamics, and this allowed the two systems to be tuned independently [4]. Moreover, because the (secondary) motor controller can be tuned independently of the primary controller, the cascaded architecture allows us to modify and iterate on the motor hardware without costly and time-consuming cold-flow testing.

The secondary controller is a proportional-integral-derivative (PID) controller [4]. Using measurements from an encoder fitted to the motor shaft, the controller outputs a motor command signal $u_2$ to reach and maintain the valve angle at the setpoint provided by the primary controller. Specifically,

$$u_2 = K_{2p} e_2 + K_{2i} \int e_2 dt + K_{2d} \frac{de_2}{dt} \quad (1)$$

where $K_{2p}$, $K_{2i}$, $K_{2d}$ are gains for the proportional, integral, and derivative terms, and $e_2$ is the deviation $u_1 - \theta$ of the valve angle $\theta$ from the setpoint $u_1$.

While the standard PID controller is well-suited for linear time-independent systems (such as the motor), the pressurisation system is highly non-linear and time varying. During a flow, the pressurant tank drops from approximately 310 bar to 42 bar, while ullage volume in the propellant tanks increases from approximately 5% to 100%. Compounding these are the effects of cooling as the gas expands from the pressurant tank, and the effect of pressurant collapse, which contribute non-linearities to the system. In addition, there is also significant non-linearity in the flow coefficient of the full port ball valves used for pressurant control and throttling. When combined, these factors present a significant challenge to accurate pressure regulation — when using a standard PID controller, we found that the system exhibits large-amplitude (>7 bar) oscillations in downstream pressure at the start of the flow due to low ullage volume, while struggling to maintain the pressure setpoint towards the end of the flow as ullage volume increases.

To tackle these challenges, we made the following modifications to the standard PID controller for the primary control loop.

First, we added a feedforward term to the controller output, which was derived using a simple model of the system. Specifically, we modelled the flow coefficient $C_v$ of the valves as a piecewise linear function of the valve angle $\theta$, $C_v(\theta) = max(0, (\alpha(\theta - \theta_0)))$, where $\alpha$ and $\theta_0$ are empirically determined (See Section 2.5). For tank eRegs, where the pressure drop across the valve is high (typically >100 bar), we assume choked flow through the valve orifice. In this regime the mass flow rate $Q$ of gaseous pressurant across the valve is given by $Q = kC_v p_p$, where $k$ is a constant, and $p_p$ is upstream (pressurant tank) pressure. For injector eRegs, through which liquid propellant flows, the volumetric flow rate across the valve is instead given by $Q = C_v \sqrt{\Delta p / \rho}$, where $\rho$ is fluid density, and $\Delta p$ is the pressure drop across the valve orifice [7]. The gases in both pressurant and propellant tanks are assumed to be ideal and at constant temperature. Combining these equations, and the ideal gas law gives us the following equations for the feedforward term in the tank and injector eRegs (Equation 2 and 3 respectively), where the pressurant and propellant tank pressures $p_p$, $p_T$ are provided by sensor measurements during the flow, $s_T$, $s_I$ are tank and injector pressure setpoints, $\gamma$ is a constant derived from empirical measurements, and $Q$ is an estimate of nominal flow rate at the injector.

$$f_{tank} = \gamma \cdot min(1, s_T/p_p) + \theta_0 \quad (2)$$

$$f_{injector} = \frac{Q\sqrt{\rho/(p_T - s_I)}}{\alpha} + \theta_0 \quad (3)$$

Adding a feedforward term helps mitigate the non-linearities in the system, since the feedback term only needs to correct for errors in the feedforward term. Despite the simplicity of our system model, we find that the feedforward term is reasonably accurate. Preliminary testing shows that even without feedback control (ie. purely feedforward), downstream pressures deviate slowly from the desired setpoint, as shown in Figure 9.

We also dynamically updated the PID controller gains during the flow, since a static set of PID gains is unable to perform well throughout the wide range of system characteristics during a flow. In particular, the small ullage volume at the start of the flow causes the system to be extremely sensitive to changes in flow rate, with the system becoming much less sensitive as the ullage volume increases throughout the flow. We therefore computed the PID controller gains dynamically using the following equations, where $k_p$, $k_i$, $k_d$, $T$ are constants to be tuned.

$$\lambda(t) = min(1, t/T)$$
$$K_{1p}(t) = \lambda(t) \cdot k_p$$
$$K_{1i}(t) = \lambda(t) \cdot k_i$$
$$K_{1d}(t) = \lambda(t) \cdot k_d \quad (4)$$

This ramps up the PID gains at the start of the flow, resulting in a less aggressive controller at low ullage volumes, while maintaining accurate control towards the end of the flow.

To throttle the engine, we maintain constant propellant tank pressures, while updating the pressure setpoints of the injector eRegs according to a predetermined profile. The pressure setpoints are chosen such that the OF (Oxidiser to Fuel) mass ratio in the combustion chamber remains constant at approximately 2.3.

*2.5 Water Flow Testing*

Preliminary tests on tank pressure control were completed using a water flow test stand to gain an understanding of the system dynamics and complete characterization of the ball valve flow coefficient for the purpose of the feedforward model. The waterflow test stand allowed for a rapid iteration cycle at a low cost without the hazards of cryogenic testing. Water was loaded into the test stand using a pump, and high pressure compressed air was obtained with a SCUBA tank pump.

For tests without injector pressure control, tank pressures were set after determining nominal pressure drops in the system to obtain the correct mass flow rates into the combustion chamber. A combination of water flows and liquid nitrogen / propane cold flows were utilised to collect data and the losses in the feed system were modelled using the Darcy-Weisbach equation.

*2.6 Cold Flow Testing*

Cold flow testing was used to further verify the accuracy of our regulation hardware under more challenging conditions than water flow tests presented, and to develop a more accurate system model. A liquid bi-propellant test stand used previously for engine testing, shown in Figure 6, was modified to accommodate the eReg hardware and associated avionics systems.

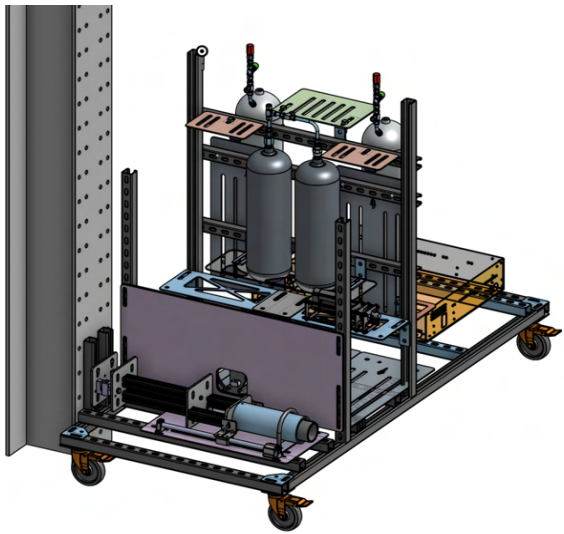

**Fig. 6.** Test stand used for cold flow and static fires

We conducted cold flow testing in two stages: we first tested the system without injector eRegs (ie. tank eRegs only) to verify the hardware design and basic functionality, then ran throttling tests with both tank and injector eRegs. Initial tests were performed with liquid nitrogen in both propellant tanks, while acceptance testing was carried out with liquid nitrogen and propane. Figure 7 shows a picture of our test stand in operation.

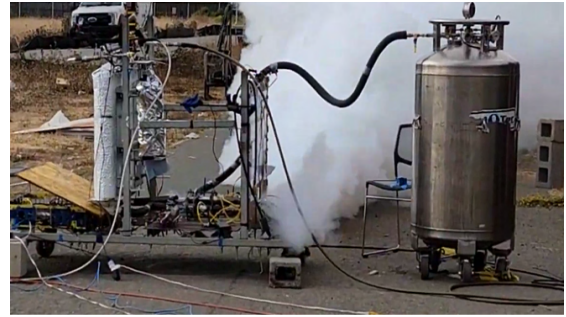

**Fig. 7.** Cold flow test stand in operation

For throttling tests, we used the following profile for injector pressure setpoints: first, we ramp up to initial pressure $p_1$, which we hold constant for time $t_1$ before throttling up to the maximum pressure $p_2$ held constant constant for time $t_2$, followed by throttling down to the minimum pressure $p_3$, which is held for the remainder of the flow.

*2.6.1 Nominal Flow Rate Testing*

During cold flow testing with nominal injector geometry, the lack of back pressure from the combustion chamber led to much higher flow rates than would be expected during static fire or flight. We found that such flow rates led to unexpectedly high pressure drops across the injector eReg valve, and a loss of control authority. Therefore, to simulate hotfire conditions more closely, "mock" injector elements were sized to permit nominal flow rates at nominal pressures, allowing tuning of the controller to be carried out with less difficulty.

Mock injector elements were designed by drilling holes at a precise orifice diameter to ensure the mass flow rate from nominal tank or injector pressures to the atmospheric pressure matched hotfire flow rates. Initial orifice release coefficients were sized based on literature, and subsequently adjusted based on performance from further testing.

*2.7 Static Fire Testing*

To verify the capabilities of our system before flight, we performed two LOx / propane static fire tests. The first test was conducted without injector eRegs, and verified the ability of the upstream regulators to maintain a constant tank pressure, while the second test demonstrated the ability to throttle the engine through a range of thrusts, and utilised both tank and injector eRegs. We used the same throttling setpoints as in our cold flow tests. A picture of the test stand is shown in Figure 12.

## 3. Results .
*3.1 Water Flow Tests*

Initial waterflow results were successful at demonstrating the potential electronic regulation system, and as a verification of our feedforward model for tank pressurisation.

In order to derive the parameters used in our feedforward model, we first performed tests to characterise the flow coefficient of our valve. The results are shown in Figure 8.

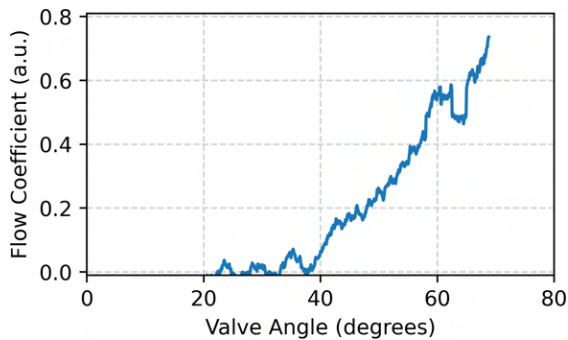

**Fig. 8.** Empirically determined flow coefficient as function of valve angle.

We then tested an open-loop, feedforward-only controller to validate our model. The results, shown in Figure 9, show that while simple, the feedforward model is accurate enough to mitigate the nonlinearities in the system.

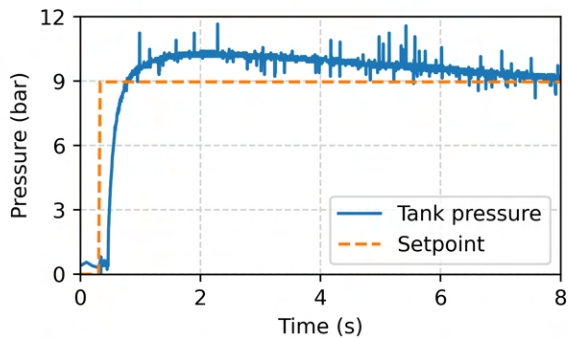

**Fig. 9.** Tank pressure regulation with feedback disabled (feedforward only) on the primary controller

However, initial flow tests with both feedforward and feedback control performed poorly due to the lack of dynamic PID and the ramp in the setpoint of the tank pressures at the beginning of flow. This forced us to compromise between an initial overshoot and poor regulation near the end of flow as the high pressure gas supply dropped in pressure and the ullage volume increased. Despite these issues, the results indicated lower steady state error than the dome regulators used as our baseline comparison test and a dramatic improvement in the amount of time needed to characterise the system after changes were made to hardware or to flow parameters.

*3.2 Cold Flow Tests*

Initial cold flow testing was used to determine the ability to regulate the propellant pressure under the demanding conditions of an injector with orifice sizing based on hotfire data, and at pressures used during hotfire. This caused an approximate mass flow rate increase of fuel, oxidiser, and pressurant of about 1.8 times due to the lack of chamber pressure arising from combustion. During early testing, the system struggled to compensate for the more rapid change in ullage volume, however aggressive gains caused overshoot when ullage volume was minimal. Tuning was made more difficult by the various rates of boiloff causing different amounts of ullage volume in our tanks, as propellant was vented to maintain a safe working pressure. The reliability of our fill level sensing hardware was low, so data from these sensors could not be incorporated into our testing or controller.

However, the use of dynamic PID gains (see Section 2.4) was effective in mitigating this issue by allowing the controller to make subtle adjustments towards the start of the flow, and make increasingly aggressive corrections as ullage volume increased. The regulated and setpoint pressures for all 4 eRegs during acceptance testing are shown in Figures 10 and 11. Besides a small initial transient, which we attribute to the propellants rapidly boiling due to a non-chilled system at the start of the test, we see that injector pressure control is accurate to within ~0.5 bar.

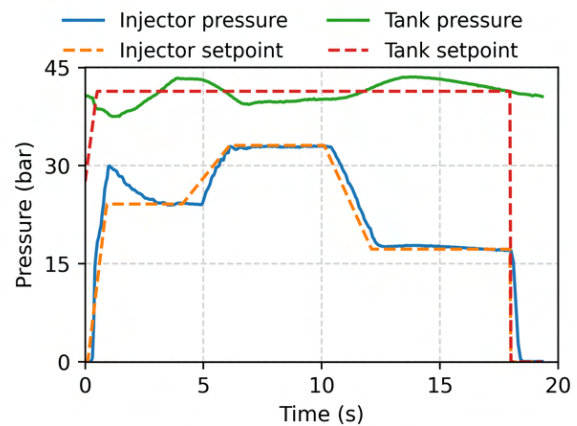

**Fig. 10.** Regulated and setpoint pressures in oxidiser tank and injector during cold flow.

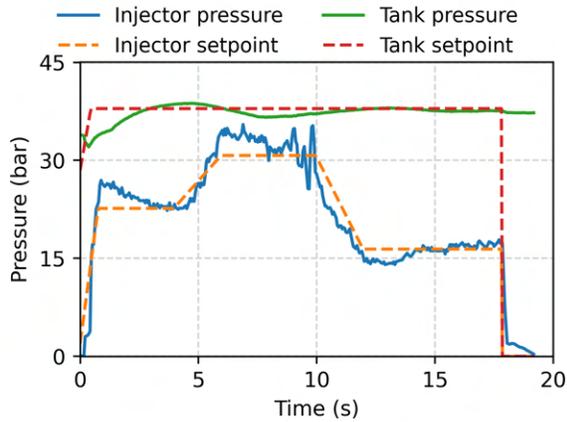

**Fig. 11.** Regulated and setpoint pressures in fuel tank and injector during cold flow.

To reduce the difficulty of controller tuning, nominal flow rate tests were performed in two different ways. During tests with an injector size for hotfire, the target tank pressures were reduced to accommodate a nominal mass flow rate across the injector. This allowed the testing team to gain more experience with the hotfire injector hardware, and perform associating testing. Although this testing simulated somewhat nominal regulation, the regulation hardware was not pressurised to MEOP, and thus was not satisfactory for acceptance testing.

MEOP Flows with an undersized mock injector allowed for the most nominal flow rates and pressures for the system, allowing the control algorithm to be tuned specifically for a hotfire while putting the regulator hardware through its paces.

*3.3 Static Fire Tests*

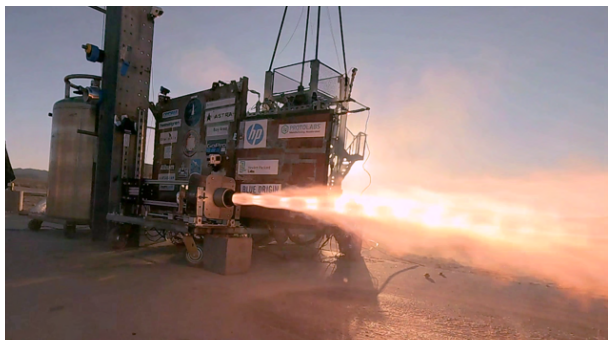

**Fig. 12.** Test stand undergoing static fire test.

We conducted a static fire to test the ability of the injector eRegs to throttle the thrust chamber up and down by limiting the mass flow rate of propellant. The static fire began with a slow ramp up in pressure to ensure smooth start up, steady state, another ramp up, steady state, a ramp down, and an additional period of steady state until propellants were exhausted 14 seconds into the burn. Figure 13 shows the regulated pressures in the oxidiser tank and injector, while Figure 14 shows the same for the fuel tank and injector.

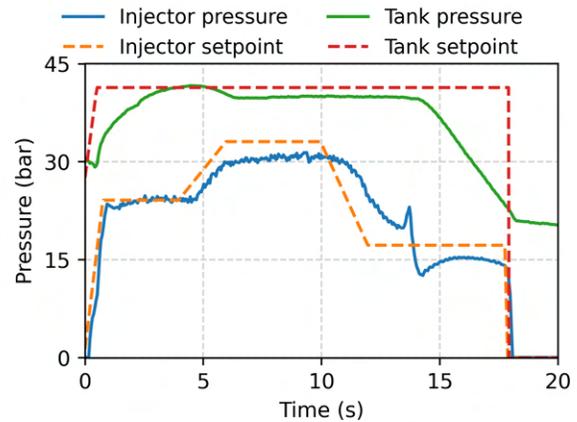

**Fig 13** Oxidizer pressures during static fire.

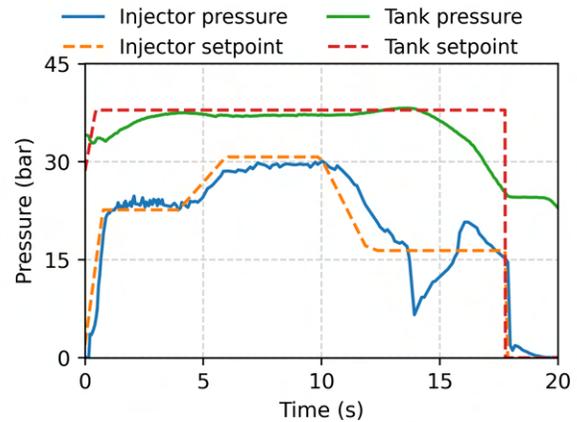

**Fig 14** Fuel pressures during static fire.

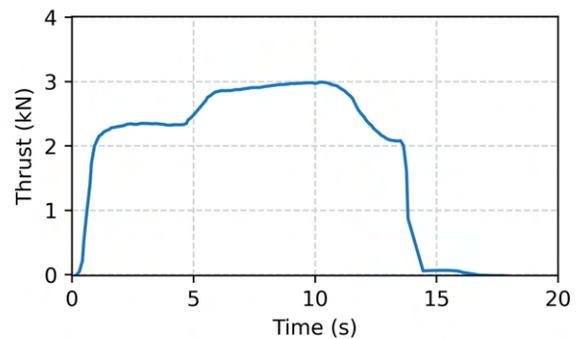

**Fig 15** Thrust curve during static fire. Our engine successfully throttled up to 3kN, then down to 2.1kN.

We see that pressure regulation for all 4 eRegs are reasonably accurate: tank pressures are within 0.5 bar of the setpoint, even when throttling, while injector pressures are within 1 bar throughout most of the burn. Note that the abrupt pressure changes at approximately 14 seconds are the result of the propane running out and ending the burn prematurely.

The controlled changes in injector pressure were able to successfully throttle the engine, as shown by the thrust curve in Figure 15. The engine achieved a maximum thrust of 3kN, and throttled down to 2.1kN before the end of the burn, achieving a 70% throttle level.

## 4. Discussion

While the control algorithm performed well during the test, we believe that significant improvements can be made to the responsiveness and accuracy of the injector pressure regulation. Specifically, the results from this first test can be used to further tune the PID controller gains. In future static fires, it would be ideal to increase the integral and proportional gains in the primary controller. This is a promising direction for future work, given that key applications of throttling such as propulsive landing require accurate and responsive control.

Although a 70% throttle level was achieved in our work, demonstration of lower throttle levels would be another direction for future work. The limit to deep throttling is generally regarded as combustion instability resulting from poor injector stiffness [1]. At low injector pressures, the pressure drop across the injector is low compared to the chamber pressure, causing poor injector performance that manifests as poor atomization and mixing, The drop in combustion performance and inconsistent injection causes instability that can cause damage to the thrust chamber or external combustion. An analysis of the theoretical limits of throttling with the current thrust chamber injector geometry and hotfire characterization could be completed to prove this analysis.

In addition, the hardware demonstrated the effectiveness of utilising full port ball valves to effectively regulate gaseous and liquid propellants, in contrast to other valves more commonly used for throttling such as v-port or needle valves [2].

## 5. Conclusions

The cumulative results from the waterflow, coldflow, and static fire campaign demonstrated the potential of electromechanical regulation of a dynamic system, opening up opportunities for small scale liquid rockets to incorporate throttling and baseline more accurate feed system pressures in their flight profiles. More accurate feed system pressures allow for a more precise OF ratio, which in turn increases the potential ISP of the engine. Further, throttling allows for altitude gains, as a more fine balance between gravity losses and drag losses can be achieved.

## Acknowledgements

We would like to thank the team at Space Enterprise at Berkeley for their support. In particular, we extend our gratitude for their help in building and operating the test infrastructure, as well as development of the electronics for this project.